\documentstyle[aps,pre,eqsecnum,multicol]{revtex}

\newcommand {\Rm} {R_{\rm max}}
\newcommand {\Rb} {R_{\rm back}}
\newcommand {\Pm} {\Pi_{\rm max}}

\begin{document}
\draft

\title{On Turbulence of Polymer Solutions}

\author{E.~Balkovsky$^{a,b}$, A.~Fouxon$^c$, and V.~Lebedev$^{c,d}$}

\address{$^{a}$ The James Franck Institute,
  University of Chicago, 5640 S. Ellis Ave., Chicago, IL,60637 \\
  $^{b}$ Department of Mathematics, University of Chicago, 5734 S.
  University Avenue Chicago, IL, 60637\\
  $^c$ Physics Department, Weizmann Institute of Science,
  Rehovot, 76100, Israel \\
  $^d$ Landau Inst. for Theoretical Physics, Moscow, Kosygina 2,
  117940, Russia}

\date{\today}

\maketitle

\begin{abstract}
  
  We investigate high-Reynolds number turbulence in dilute polymer
  solutions. We show the existence of a critical value of the Reynolds
  number which separates two different regimes. In the first regime,
  below the transition, the influence of the polymer molecules on the
  flow is negligible and they can be regarded as passively embedded in
  the flow. This case admits a detailed investigation of the
  statistics of the polymer elongations. The second state is realized
  when the Reynolds number is larger than the critical value. This
  regime is characterized by the strong back reaction of polymers on
  the flow. We establish some properties of the statistics of the
  stress and velocity in this regime and discuss its relation to the
  drag reduction phenomenon.

\end{abstract}


\begin{multicols}{2}

\section{Introduction}
  
In this paper we present a theoretical investigation of turbulence in
dilute polymer solutions. As opposed to Newtonian fluids, such
solutions possess additional macroscopic degrees of freedom related to
the elasticity of the polymer molecules. Relaxation times of elastic
stresses can be comparable with time-scales of the flow which means
that the relation between the stress and velocity gradient is
non-local. Due to the non-trivial interaction of inertial and elastic
degrees of freedom, the polymer solutions exhibit a variety of
regimes. For example, a turbulent-like state has been recently
observed at very low Reynolds' numbers \cite{00GS}. Here we will
consider the more familiar situation of turbulence at high Reynolds
numbers. Probably, the most striking effect of polymers on the high
Reynolds number flows is the drag reduction phenomenon. The addition
of long-chain polymers in concentrations as small as $10^{-5}$ by
weight can induce a substantial reduction of the drag force needed to
push a turbulent fluid through a pipe \cite{Virk,McComb,95GB}.
  
The reason why small amounts of polymer can significantly modify
properties of the fluid is the flexibility of polymer molecules. At
equilibrium a polymer molecule coils up into a spongy ball of a radius
$R_0$. The value of $R_0$ depends on the number of monomers in the
molecule, which is usually very large. For a dilute solution with the
concentration, $n$, satisfying $nR_0^3\ll1$, the influence of
equilibrium size molecules on the hydrodynamic properties of the fluid
can be neglected. When placed in a non-homogeneous flow, such a
molecule is deformed into an elongated structure that can be
characterized by its end-to-end distance $R$. If the number of
monomers in a typical polymer molecule is large, the elongation $R$
can be much larger than $R_0$. The influence of the molecules on the
flow increases with their elongation and may become substantial when
$R\gg R_0$.

The deformation of the molecule is determined by two processes, the
stretching by the velocity gradients and the relaxation due to the
elasticity of the molecule. To understand how a molecule resists the
deformation by the flow, let us consider its relaxation. Recent
experiments with DNA molecules indicate that the relaxation is linear
in the wide region of scales $R_0\ll R\ll \Rm$, where $\Rm$ is the
maximum extension \cite{Chu1}. In the case of polymer molecules,
theoretical arguments and numerics presented in \cite{99HQ} support
the linear relaxation. These results can be understood if we assume
that at $R\gg R_0$ the role of excluded volume and hydrodynamic
interactions between the monomers becomes negligible. Then the random
walk arguments suggest that the entropy of polymer molecules is
quadratic in the range $R_0\ll R\ll \Rm$ implying linear relaxation.
Whether the polymers are excited by the flow is determined by the
softest relaxation mode that describes the dynamics of the elongation
$R$. In the absence of stretching, the relaxation of $R$ is described
by $\dot{R}=-R/\tau$, where $\tau$ is the largest of the polymer
relaxation times. The relaxation time is $R$-independent at $R_0\ll
R\ll \Rm$. If the end-to-end distance $R$ is of the order of the
maximum extension, $\tau$ starts to depend on $R$ and the dynamics of
the molecule becomes nonlinear.

The behavior of the molecule in an inhomogeneous steady flow depends
on the value of the Weissenberg number, ${\rm Wi}$, defined as the
product of the characteristic velocity gradient and $\tau$. When a
polymer molecule is placed in a flow, smooth at the scale $R$, the
velocity difference between the end-points is proportional to $R$
multiplied by the characteristic value of velocity gradient. At
${\rm~Wi}\ll 1$ relaxation is fast compared to the stretching time and
the polymer always relaxes to the equilibrium size, $R_0$. The
behavior of the polymer at ${\rm Wi}\gtrsim 1$ depends on the geometry
of the flow. For purely elongational flows the molecule gets aligned
along the principal stretching direction. If the velocity gradient is
larger than the inverse relaxation time, i.e. ${\rm Wi}\gtrsim 1$, the
elastic response becomes too slow in comparison with the stretching
and the molecule gets substantially elongated \cite{Lumley73}. The
sharp transition from the coiled state to the strongly extended state
is called the coil-stretch transition. Rotation can suppress the
transition and even damp it completely since the molecule does not
always point in the stretching direction (see e.g. \cite{Lumley72}).
For example, no coil-stretch transition occurs in the case of a shear
flow, which is a combination of elongational and rotational flows.

In contrast to the steady flows, a polymer molecule moving in a smooth
random flow alternately enters regions of high and low stretching. As
the intensity of the flow increases the effect of the stretching
becomes more pronounced. One can generally assert the existence of the
coil-stretch transition. This has been first demonstrated by Lumley
\cite{Lumley72}, who considered the situation where the characteristic
time of variations of velocity gradient is much larger than the
inverse of the characteristic value of the gradient. He showed that if
the amplitude of velocity gradient fluctuations is large enough, the
expectation value of $R^2$ grows with time, which signifies the
coil-stretch transition. We have demonstrated in \cite{short} that the
coil-stretch transition occurs in any random flow and established a
general criterion for the transition. In particular, the transition
occurs in the situation where the time of velocity gradient variation
is of the order of the inverse of its characteristic value, which is
likely to be the case for real flows. The coil-stretch transition in
random flows is controlled by the parameter $\lambda_1\tau$, where
$\lambda_1$ is the average logarithmic divergence rate of nearby
Lagrangian trajectories, to be referred to as the principal Lyapunov
exponent. It is positive for an incompressible flow
\cite{Furst,Zeldovich}. The molecules are weakly stretched if
$\lambda_1\tau<1$ and strongly stretched otherwise. Therefore for
random flows the parameter $\lambda_1\tau$ plays the role of the
Weissenberg number.

To describe the behavior of a polymer molecule in turbulent flows, let
us briefly review the basic properties of turbulence of incompressible
Newtonian fluids. A high Reynolds number flow consists of chaotic
motions from a wide interval of scales, $\eta\ll r\ll L$, where $L$ is
the scale at which the flow is excited and $\eta$ is the viscous
scale. The energy pumped at the scale $L$ cascades down to the scale
$\eta$, where it is dissipated. The size of polymer molecules is
usually much smaller than the viscous scale. Viscosity makes the flow
smooth at $r\ll \eta$, i.e. the velocity difference between two points
is given by the velocity gradient multiplied by the distance. Then,
the stretching of molecules is determined by the gradient of velocity,
which should be considered random in a turbulent flow. The Lyapunov
exponent can be estimated as the characteristic value of the velocity
gradient, which is determined by the eddies at the viscous scale. As
Reynolds number increases, the velocity gradient increases, and so
does $\lambda_1\tau$. At some value of Reynolds number it reaches the
value $1$ and the coil-stretch transition occurs.

Several mechanisms can limit the stretching of polymers. The first one
is the internal non-linearity of the elasticity of the polymer
molecules. If this mechanism dominates, then above the transition the
molecules are stretched up to the maximal elongation $\Rm$. An
alternative explanation has been proposed by Tabor and de Gennes
\cite{Tabor}. It is based on the fact that if the elongation of a
polymer molecule is larger than the viscous length of turbulence,
$\eta$, the elastic force always wins over the stretching. Estimates
using the parameters of typical polymer solutions show that this
situation is difficult to realize. Therefore, we will assume that the
inequality $R\ll\eta$ is satisfied. It will also enable us to write
local equations describing the dynamics of elastic stresses. Another
mechanism is the back reaction of the polymers on the flow. It is
caused by the collective contribution of coherently deformed polymer
molecules into the stress tensor. This elastic part of the stress
grows with the elongation of the molecules. When it becomes of the
order of the viscous stresses existing in the flow, the polymers
modify the flow around them and the stretching diminishes. As a
result, a dynamic equilibrium is realized at the characteristic
elongation, $\Rb$. The total polymer stress is proportional to $nR^2$,
so that $\Rb$ depends on the polymer concentration $n$. Therefore if
the concentration is large enough, the value of $\Rb$ is much smaller
than $\Rm$. We will consider the effect of the back reaction under the
assumption $\Rb\ll\Rm$. Probably, the condition $\Rb\ll\Rm$ is
necessary for the existence of a stationary state, because the polymer
molecules stretched up to $\Rm$ are intensively destroyed by the flow.

Above the coil-stretch transition the back reaction modifies the
small-scale properties of turbulent flows, which leads to the
emergence of a new scale, $r_*>\eta$. Large-scale eddies with the
sizes $r>r_*$ do not excite elastic degrees of freedom, so that the
usual inertial energy cascade is realized at these scales. At smaller
scales inertial and elastic degrees of freedom exchange energy, which
is dissipated mainly due to the polymer relaxation. The energy cascade
terminates at $r_*$, so that $r_*$ plays the role of a new dissipation
scale.

The plan of the paper is as follows. In Sec. \ref{basiceq} we
introduce a system of equations describing the coupled dynamics of
inertial and elastic degrees of freedom. In Sec. \ref{pastensor} we
study the situation when the back reaction of polymers is small and
can be disregarded. We find the probability distribution function of
the elastic stress tensor and examine its correlation functions. In
Sec. \ref{saturation} we study the influence of the back reaction on
the flow and establish some properties of the velocity and stress
statistics. In Conclusion we summarize our results and discuss their
implications for the drag reduction phenomenon. In appendix
\ref{integr} we present a detailed derivation of the probability
density function from Sec. \ref{pastensor}. Appendix \ref{model} is
devoted to a simple model illustrating some aspects of the interaction
between the flow and polymers. Preliminary results of this work have
been published in \cite{short}.

\section{Basic Relations}
\label{basiceq}

Following Hinch \cite{Hinch77} let us consider the dynamics of a
polymer molecule in a smooth velocity field. The degree of freedom
related to the elongation of the molecule is described by the vector
${\bbox R}$, connecting the end-points of the molecule. The equation
describing the dynamics of ${\bbox R}$ in the absence of a surrounding
flow is
\begin{eqnarray}&&
\partial_t R_i+\Gamma\frac{\partial E}{\partial R_i}=\zeta_i \,,
\label{tlhs0} \end{eqnarray}
where $E$ is the free energy of the molecule, $\zeta_i$ is the thermal
noise, and $\Gamma$ is the kinetic coefficient which determines the
relaxation of the molecule. The correlation function of $\zeta$ is
\begin{eqnarray}&&    
\langle\zeta_i(t)\zeta_j(t')\rangle
=2k_BT\,\Gamma\delta_{ij}\delta(t-t') \,,
\label{tlhs1}\end{eqnarray} 
where $T$ is temperature and $k_B$ is the Boltzmann constant. If the
size of molecules is much smaller than the viscous length, which we
assume, the molecule moves in a constant gradient flow. Its influence
is described by the equation
\begin{eqnarray}&&
\partial_t R_i-R_j\nabla_j v_i
+\Gamma\frac{\partial E}{\partial R_i}=\zeta_i \,,
\label{tlhs} \end{eqnarray}
where the stretching term $-R_j\nabla_j v_i$ is added to Eq.
(\ref{tlhs0}) The velocity derivative must be evaluated at the
position of the molecule. To avoid misunderstanding note that we mean
the flow which is ``external'' to the molecule, excluding the velocity
induced by the relative motions of its chains.

The entropy of the molecule has a quadratic dependence on the
elongation, $R$, in a wide interval \cite{Kit}. It implies that the
molecule can be treated in terms of elasticity theory with a Hook
modulus $K_0$, so that $E=K_0 R^2/2$. This expression is correct
provided $R\ll \Rm$, where $\Rm$ is the maximum elongation of the
molecule. The equilibrium size of the molecule, $R_0$, can be
estimated from the condition $E\sim k_BT$ as $\sqrt{k_B T/K_0}$.
Substituting the energy into Eq. (\ref{tlhs}) we get
\begin{eqnarray}&&
\partial_t R_i-R_j\nabla_j v_i
+R_i/\tau=\zeta_i \,,
\label{tlhs3}\end{eqnarray}
where $\tau=(\Gamma K_0)^{-1}$. We see that $\tau$ is the molecular
relaxation time.

Generally, the kinetic coefficient $\Gamma$ or the relaxation time
$\tau$ is a function of $R$, which reflects non-linear character of
the molecule relaxation \cite{87BCAH}, related to such effects as
internal hydrodynamic interaction of chains in the polymer molecule.
For example, the finitely extendible nonlinear elastic model
\cite{87BCAH} assumes $\tau\propto 1-{\sl R}^2/\Rm^2$. One expects
that a dependence of $\Gamma$ and $\tau$ on $R$ can be disregarded at
$R\ll \Rm$. Below we study this situation. Possible statistical
consequences of the non-Hookean dependence of the free energy on $R$
have been investigated in \cite{misha}.

Equations (\ref{tlhs},\ref{tlhs3}) assume that $R$ is the only mode
related to the molecular deformation, which is an idealization.
Actually, the molecule has many deformational degrees of freedom that
have different relaxation times. They have been observed
experimentally \cite{Chu1}. Nevertheless, in the turbulent flows, only
the mode with the largest relaxation time is strongly excited whereas
other modes are excited at most weakly. Thus, Eq.  (\ref{tlhs3})
should be considered as the equation describing the principal mode.

\subsection{Continuous media equations}
\label{conti}

To study the dynamics at the scales much larger than the inter-polymer
distance the polymer solution can be regarded as a continuous medium.
The appropriate description is done in terms of macroscopic quantities
which are averages of microscopic variables over the volume. The
polymer molecules are characterized by the average conformation tensor
\begin{eqnarray}&& {\sl A}_{ij}=\langle R_i R_j \rangle \,.
\label{tlhs2} \end{eqnarray} 
The volume of averaging should contain a large number of polymer
molecules and be smaller than the characteristic scales of the
processes under consideration. The tensor ${\sl A}$ can also be
interpreted as the average over the statistics of the thermal noise
${\bbox\zeta}$. Equation for ${\sl A}_{ik}$ follows from Eqs.
(\ref{tlhs1},\ref{tlhs3})
\begin{eqnarray}&&
\partial_t {\sl A}_{ij}+(\bbox v\nabla){\sl A}_{ij}
={\sl A}_{kj}\nabla_k v_i+{\sl A}_{ik}\nabla_k v_j\nonumber\\&&
-\frac{2}{\tau}\left({\sl A}_{ij}-{\sl A}_0\delta_{ij}\right)\,,
\label{intro2}\end{eqnarray}  
where ${\sl A}_0=k_BT/K_0$. Equation (\ref{intro2}) is linear in $A$,
which is correct provided $A\ll \Rm^2$.

Equation (\ref{intro2}) should be supplemented by the equation for the
fluid velocity. This equation is a consequence of the momentum
conservation law. In order to derive it one should take into account
the contribution of the inner elastic forces of polymer molecules to
the total stress tensor of the fluid. If $\Pi_{ik}$ is the elastic
stress tensor per unit mass, the polymer contribution is
$\varrho\Pi_{ik}$ where $\varrho$ is the mass density of the fluid.
In the Hookean approximation
\begin{eqnarray}&&
\Pi_{ik}=\frac{K_0 n}{\varrho}
{\sl A}_{ik}-\Pi_0\delta_{ik} \,,
\label{stress} \end{eqnarray}
where $\Pi_0=K_0{\sl A}_0n/\varrho=(n/\varrho)k_BT$ originates from
the thermal noise $\bbox\zeta$ in Eq. (\ref{tlhs3}). Here $n$ is the
concentration of the polymer molecules. If the flow is incompressible,
i.e. $\varrho$ is a constant and $\nabla\cdot{\bbox v}=0$, the momentum
conservation law reads
\begin{eqnarray}&&
\partial_t v_i+(\bbox v\nabla)v_i
+\varrho^{-1}\nabla_i P=\nu \nabla^2 v_i
+\nabla_k \Pi_{ik}+f_i \,.
\label{intro1} \end{eqnarray}
Here $P$ is the pressure, $\nu$ is the kinematic viscosity of the
solvent and ${\bbox f}$ is the external force (per unit mass) driving
the flow. Equation (\ref{intro1}) is a generalization of the
Navier-Stokes equation to the case of viscoelastic fluids. To simplify
the consideration we assume that ${\bbox f}$ is homogeneously
distributed over space. It is a common belief that this case does not
differ qualitatively from that realized for real experimental setups,
where pumping is usually related to the boundaries.

The applicability condition of Eq. (\ref{intro2}) is $A\ll \Rm^2$.  It
can be rewritten in terms of elastic stress tensor as
$\Pi\ll\Pm\equiv{K_0 n}{\varrho}^{-1}\Rm^2$. We assume that this
condition is satisfied for relevant fluctuations. The interaction
between ${\bbox v}$ and $\Pi$ turns on if $\Pi$ exceeds the viscous
stress, $\nu\nabla{\bbox v}$. The latter can be estimated as
$\nu\lambda_1$ where $\lambda_1$ is the average logarithmic divergence
rate of nearby Lagrangian trajectories. Under the condition
$\Pm\ll\nu\lambda_1$ the polymer molecules exert no influence on the
flow except for a small renormalization of the viscosity of the
solution. Thus, the inequality $\Pm\gg\nu\lambda_1$ is a necessary
condition for the polymers to have non-trivial effects.

The free energy of the viscoelastic fluid is the sum of the kinetic
and elastic contributions
\begin{eqnarray}&&
{\cal F}=\int\!  d{\bbox r}\,
\left\{\frac{\varrho}{2}v^2+\frac{n K_0}{2}\left[
{\rm tr}\,{\sl A}-{\sl A}_0\ln({\rm det}\,{\sl A}/{\sl A}_0)
\right]\right\} \,,
\nonumber
\end{eqnarray}
where the second term represents the entropy of the molecules. Then we
find from Eqs.  (\ref{intro2}-\ref{intro1})
\begin{eqnarray}&&
\frac{\partial {\cal F}}{\partial t}
=\varrho\int\!  d{\bbox r}\,
{\bbox f}\cdot {\bbox v}\nonumber\\&&
-\varrho\int\!  d{\bbox r}\,\left\{\nu(\nabla_i v_j)^2
+\frac{1}{\tau}{\rm tr}\,\left[\Pi
\left(\Pi_0+\Pi\right)^{-1}\Pi\right]\right\} \,.
\label{energy1} 
\end{eqnarray}
This equation provides a mathematical formulation of the energy
balance: the force supplies the energy, which is then dissipated due
to viscosity and relaxation of polymers. Note that the second integral
in Eq. (\ref{energy1}) has a definite sign, as it should be for a
dissipative term. Relative contributions of the viscous and the
elastic terms to the energy dissipation can be different.

If the forcing is statistically homogeneous, then statistically
homogeneous steady state is realized. It can be described in terms of
correlation functions of ${\bbox v}$ and ${\sl A}$ (or ${\bbox v}$ and
$\Pi$), which are averages over the statistics of the pumping force
${\bbox f}$ or over space. In the steady state the average value of
${\partial}{\cal F}/{\partial t}$ is equal to zero.  Therefore we get
from Eq.  (\ref{energy1})
\begin{eqnarray}&&
\nu \langle (\nabla_i v_j)^2\rangle
+\frac{1}{\tau}\left\langle {\rm tr}\,
\left[\Pi\,(\Pi+\Pi_0)^{-1}\Pi\right]
\right\rangle=\epsilon \,, 
\label{dissip} \end{eqnarray}
where $\epsilon=\langle {\bbox f} {\bbox v} \rangle$ is the mean
energy injection rate (per unit mass) by the external force.

Generally, the diffusion term $\kappa\nabla^2{\sl A}$ should also be
added to the right-hand side of Eq. (\ref{intro2}). The diffusion
coefficient, $\kappa$, is small due to a large number of monomers.  It
is possible to show that the limit $\kappa\to0$ is regular, so we can
disregard the diffusion. The diffusion term can play a minor role for
scales $r\lesssim \sqrt{\kappa/\lambda_1}$. If
$\sqrt{\kappa/\lambda_1}$ is smaller than the intermolecular distance,
then the diffusivity is irrelevant in the whole region of
applicability of the macroscopic approach.

\subsection{Lagrangian Description}
\label{lagrange}

Equation (\ref{intro2}) can formally be solved in the Lagrangian
reference frame. Let us introduce $\tilde{\sl A}(t,\bbox r)={\sl
A}[t,\bbox x(t,\bbox r)]$, where $\bbox x(t,\bbox r)$ is the
Lagrangian trajectory defined by the relations
\begin{eqnarray}&&
  \partial_t {\bbox x}={\bbox v}(t,{\bbox x})\,,\quad 
 {\bbox x}(t_0,{\bbox r})={\bbox r} \,.
\label{lagrtr} \end{eqnarray}
The condition ${\bbox x}(t_0, {\bbox r})={\bbox r}$ ensures that the
fields ${\sl A}$ and $\tilde{\sl A}$ coincide at $t=t_0$.  The point
${\bbox r}$ plays the role of a Lagrangian marker. The tensor
$\tilde{\sl A}(t,\bbox r)$ satisfies the matrix equation following
from Eq. (\ref{intro2})
\begin{eqnarray}&&
\partial_t\tilde{\sl A}=\sigma \tilde{\sl A}
+\tilde{\sl A}\sigma^T-\frac{2}{\tau}
\left(\tilde{\sl A}-{\sl A}_0\right),
\label{inlf} \\ &&
\sigma_{ij}(t,\bbox r)
=\nabla_j v_i[t,{\bbox x}(t,\bbox r)] \,.
\label{sigma} \end{eqnarray} 
Here $\sigma$ is the tensor of the velocity derivatives in the
Lagrangian frame and the superscript $T$ denotes a transposed matrix.
Due to the causality, the value of the field ${\sl A}(t,{\bbox r})$ is
determined by its dynamics at times $t'<t$. Therefore we will be
interested in the backward in time Lagrangian evolution of ${\sl A}$
described by Eqs. (\ref{lagrtr},\ref{inlf}).

A solution of Eq. (\ref{inlf}) can be written in terms of the
Lagrangian mapping matrix $W$ defined by the relations
\begin{eqnarray}&&
\partial_t W(t,t')=\sigma(t) W(t,t')\,, \qquad W(t', t')=1\,. 
\label{mmm1} \end{eqnarray}
The matrix $W$ is defined for a given Lagrangian trajectory, and
therefore it depends on its marker ${\bbox r}$. For brevity we omit
the argument ${\bbox r}$ in $W$.  The matrix $W$ describes the
deformation of infinitesimal fluid volumes.  For example, the
separation, $\delta{\bbox x}$, between two close Lagrangian particles
changes according to
\begin{eqnarray}&&
\delta{\bbox x}(t)=W(t,t')\,\delta{\bbox x}(t')\,.
\label{divl}
\end{eqnarray}
It follows from Eq. (\ref{divl}) that $W_{ij}(t,t_0,{\bbox r})
={\partial x_i(t,{\bbox r})}/{\partial r_j}$. The incompressibility
condition $\nabla\cdot{\bbox v}=0$ is formulated in terms of $\sigma$ as
${\rm tr}\,\sigma=0$. A consequence of Eq. (\ref{mmm1}) is
\begin{eqnarray}&&
{\rm det}\, W=1 \,.
\label{unim} \end{eqnarray}
Using Eqs. (\ref{inlf},\ref{mmm1}) we obtain
\begin{eqnarray}&&
\tilde{\sl A}(t, \bbox r)=\frac{2{\sl A}_0 e^{-2t/\tau}}{\tau}
\int_{-\infty}^t\!\!\!dt'\, W(t,t',{\bbox r}) 
W^{T}(t,t',{\bbox r})e^{2t'/\tau}\,.
\nonumber \end{eqnarray}
At $t=t_0$ this equation gives
\end{multicols}
\begin{eqnarray}&& 
{\sl A}(t_0,\bbox r)=\frac{2{\sl A}_0}{\tau}
\int^{\infty}_0 \!\!dt\, W(t_0,t_0-t,{\bbox r})
W^{T}(t_0,t_0-t,{\bbox r})e^{-2t/\tau}\,.
\label{basic} \end{eqnarray}
\begin{multicols}{2}

\noindent
It is easy to understand the meaning of Eq. (\ref{basic}). The
polymers are advected along the Lagrangian trajectories being
stretched by the velocity gradient and relaxing to their equilibrium
shape due to elasticity. The value of the conformation tensor ${\sl
A}$ is determined by the sum of the contributions of these processes
at earlier times picked along the Lagrangian trajectory arriving at
$\bbox{r}$. The term $WW^{T}$ describes the stretching and the
exponential factor accounts for relaxation.

Expression (\ref{basic}) shows that when calculating correlation
functions, the volume averages can be substituted by averages over the
statistics of $W$.

\section{Passive elastic tensor in a random velocity field}
\label{pastensor}

In this section we consider the polymer molecules as passive objects,
i.e. we assume that the inertial properties of the fluid are not
perturbed by the polymer elasticity. In other words, we assume that
the term $\nabla_k\Pi_{ik}$ in Eq. (\ref{intro1}) can be disregarded
so that the dynamics of ${\bbox v}$ is independent of the polymer
dynamics.  Equation (\ref{intro2}) determines the statistics of the
conformation tensor ${\sl A}$ provided the statistical properties of
${\bbox v}$ are known. We consider a high Reynolds number flow and
assume that its statistics is stationary, spatially homogeneous and
isotropic.

One might think that in order to examine the correlation functions of
${\sl A}$ one needs to know the precise statistics of the velocity
field described by the Navier-Stokes equation. However, we will show
that the statistics of ${\sl A}$ is universal, i.e. it does not depend
on the details of the velocity statistics. The crucial property
underlying the universality is a finite Lagrangian correlation time of
the velocity derivatives matrix $\sigma$ \cite{Pope}.

Equation (\ref{intro2}) shows that the Lagrangian dynamics of the
polymer stress tensor is determined by velocity gradients, which are
related to the viscous scale $\eta$ of the turbulence. A typical value
of the velocity gradient can be estimated as $\lambda_1$, where
$\lambda_1$ is the logarithmic divergence rate of nearby Lagrangian
trajectories. Then it follows from Eq.  (\ref{intro1}) that the
feedback of the polymers on the flow can be neglected if $\Pi\ll
\nu\lambda_1$. This is the applicability condition of the passive
approach. As we show below, the condition $\Pi\ll \nu\lambda_1$ is
satisfied for typical fluctuations if $\lambda_1\tau<1$.

A formal solution of Eq. (\ref{intro2}) is given by Eq.
(\ref{basic}). The condition $\lambda_1\tau<1$ means that the
exponentially decaying factor $\exp(-{2t}/{\tau})$ in Eq.
(\ref{basic}) dominates over the product $WW^{T}\sim\exp[2\lambda_1
t]$. In this case, for a typical velocity fluctuation the integral
over $t$ converges at $t\sim\tau$ and therefore ${\sl A}$ fluctuates
near ${\sl A}_0$. In addition to the strong peak at ${\sl A}\sim {\sl
  A}_0$, the probability distribution function of ${\sl A}$ has a
power tail. This has been demonstrated in \cite{short} in terms of the
molecule elongation ${\bbox R}$. The fluctuations ${\sl A}\gg {\sl
  A}_0$ are formed if the product $WW^{T}$ in Eq.  (\ref{basic}) is
anomalously large for a long time. Such events can be described in
terms of a universal scheme (see Ref. \cite{BF99}) shortly presented
below.

\subsection{Long-Time Lagrangian Statistics}
\label{long}

Let us briefly review the long-time statistical properties of the
Lagrangian mapping matrix $W$, determined by Eqs.
(\ref{sigma},\ref{mmm1}). We consider $W(t_+, t_-)$ at $t_+>t_-$ and
assume that $t_+-t_-$ is much larger than the Lagrangian correlation
time $\tau_{\sigma}$ of the velocity derivatives matrix (\ref{sigma}).
If the velocity statistics is homogeneous in time, the probability
distribution of $W(t_+,t_-)$ depends on the difference $t_+-t_-$ only.
Equation (\ref{mmm1}) implies that at $t_+-t_-\gg\tau_\sigma$ the
matrix $W$ is a product of a large number of independent matrices.
This is the main reason for the universality in the statistics of $W$.

It is convenient to decompose the matrix $W$ as
\begin{eqnarray}&&
W(t_+, t_-)=M\Lambda N \,, 
\label{product} \end{eqnarray}
where $\Lambda$ is a diagonal matrix, and $M$ and $N$ are orthogonal
matrices \cite{Orszag}. We denote the diagonal elements of $\Lambda$
as $e^{\rho_1}$, $e^{\rho_2}$, and $e^{\rho_3}$, and assume that they
are ordered: $\rho_1>\rho_2>\rho_3$. As a consequence of the
constraint (\ref{unim}) we have $\rho_1+\rho_2+\rho_3=0$. Equation
(\ref{mmm1}) can be rewritten in terms of $\rho_i$, and the matrices
$M$ and $N$. The equations for $\rho_i$ are
\begin{eqnarray}&&
\frac{\partial\rho_i}{\partial t_+}=\tilde\sigma_{ii}\,,
\label{rho} \end{eqnarray}
where $\tilde\sigma=M^T\sigma M$ and no summation over the repeating
index $i$ is implied.  The matrices $M$ and $N$ satisfy $\partial_t
N=\Omega_1 N$ and $\partial_t M=M\Omega_2$, where
\begin{eqnarray}&&
  \Omega_{1ik}=\frac{\tilde\sigma_{ik}+\tilde\sigma_{ki}}
  {2\sinh(\rho_i-\rho_k)}\,,\quad
  \Omega_{2ik}=\frac{\tilde\sigma_{ik}e^{2\rho_k}+\tilde\sigma_{ki}e^{2\rho_i}}
  {e^{2\rho_k}-e^{2\rho_i}}\,,
\nonumber \end{eqnarray}
for $i\not= k$ and $\Omega_{1ik}=\Omega_{2ik}=0$ for $i=k$.
It is possible to show that the eigenvalues of $W$ repel each other,
so that the
inequalities $e^{\rho_1}\gg e^{\rho_2}\gg e^{\rho_3}$ are satisfied at
$t_+-t_-\gg\tau_\sigma$ \cite{BF99}. Then the matrix $\Omega_1$ tends
to zero exponentially fast, i.e. $N$ is determined by times of the
order of $\tau_{\sigma}$ in the vicinity of $t_-$.  The matrix
$\Omega_2$ becomes $\rho$-independent at $t_+-t_-\gg\tau_\sigma$ and
the evolution of $M$ is decoupled from that of $\rho_i$. Then the
value of $M$ is determined by the time of the order of $\tau_\sigma$
at $t\approx t_+$, i.e. at $t_+-t_-\gg \tau_\sigma$ it becomes
$t_-$-independent.

The solution of Eq. (\ref{rho}) is
\begin{equation}
\rho_i=\int_{t_-}^{t_+}dt'\,
\tilde\sigma_{ii}(t') \,.
\label{sol} \end{equation}  
where the right-hand side of Eq. (\ref{sol}) is an integral of a
random process independent of $\rho_i$. Equation (\ref{sol}) shows
that the variables $\rho_i$ fluctuate around their average values
$\lambda_i(t_+-t_-)$. Here the constants $\lambda_i$ are equal to
$\left\langle\tilde\sigma_{ii}\right\rangle$. They are called the
Lyapunov exponents of the flow. Generally, the spectrum of the
Lyapunov exponents is non-degenerate: $\lambda_1>\lambda_2>\lambda_3$,
which is a necessary condition for the formalism to be
self-consistent. The incompressibility condition ensures the identity
$\lambda_1+\lambda_2+\lambda_3=0$, which implies $\lambda_1>0$ and
$\lambda_3<0$. Using relation (\ref{divl}) one can show that
$\lambda_1$ is indeed the average logarithmic rate of the divergence
of two Lagrangian trajectories:
\begin{eqnarray}&& \left\langle\frac{d\ln|{\bbox
      \delta\rho}|}{dt}\right\rangle=\lambda_1 
\nonumber \end{eqnarray}
Similarly, $\lambda_1+\lambda_2=-\lambda_3$ is the average logarithmic
rate of the area growth.

Under the condition $t_+-t_-\gg\tau_\sigma$ the quantity $\rho_i$ can
be considered as a sum of a large number of independent random
variables. It is known from the statistical mechanics (see e.g.,
\cite{Landau}) that the distribution of such quantities is given by
the exponent of an extensive function. In our case the probability
distribution function (PDF) of $\rho_i$ is
\begin{eqnarray}&& {\cal P}(t, \rho_1, \rho_2,
\rho_3)\propto \frac{1}{t} \exp\left[-tS\left(\frac{\rho_1-\lambda_1
      t}{t}, \frac{\rho_3-\lambda_3 t}{t}\right)\right] \nonumber\\&&
\times\delta(\rho_1+\rho_2+\rho_3)\,,
\label{pdf1}
\end{eqnarray} 
where $t=t_+-t_-$ and $\rho_1>\rho_2>\rho_3$ is implied \cite{BF99}.
The main exponential factor of the PDF has a self-similar form
described by the function $S$ called entropy function (see
\cite{BF99,CFKV99,Ellis}). It is positive, convex and has a minimum at
zero values of its arguments. The precise form of $S$ is determined by
details of the velocity statistics. The PDF has a sharp maximum at
$\rho_i=\lambda_i t$. In its vicinity the function $S$ has a quadratic
expansion, i.e. the distribution of $\rho$ is Gaussian.  However, if
one is interested in the expectation values of exponential functions
of $\rho_i$, they are determined by the wings of the PDF where the
Gaussian approximation is invalid. This entails the use of the whole
entropy function.

To average the functions of $\rho_1$ only, one can introduce the
reduced probability distribution function,
\begin{eqnarray}&&
{\cal P}(t,\rho_1)\propto
\frac{1}{\sqrt{t}}\exp
\left[-tS_1\left(\frac{\rho_1-\lambda_1 t}{t}\right)\right] \,,
\label{dis1} \end{eqnarray}
which is an integral of ${\cal P}(t,\rho_1,\rho_2,\rho_3)$ over $\rho_2$ 
and $\rho_3$. At small $x$ the function $S_1(x)$ can be written as
\begin{eqnarray}&&
S_1(x)\approx \frac{x^2}{2\Delta} \,.
\label{parab} \end{eqnarray}
Here $\Delta=\int\!dt\,\langle\langle\tilde
\sigma_{11}(t)\tilde\sigma_{11}(0)\rangle\rangle$ (where double
brackets designate irreducible correlation function) determines the
dispersion of $\rho_1$: $\langle(\rho_1-\lambda_1 t)^2\rangle\approx
t\Delta$. Expansion (\ref{parab}) is sufficient to describe typical
fluctuations of $\rho_1$, whereas the whole function $S$ is needed to
describe rare events.

In the passive regime the statistics of velocity gradients is
determined by the fluctuations at the viscous scale $\eta$. The
Lagrangian correlation time $\tau_{\sigma}$ is the turnover time at
this scale. It can also be estimated as $\lambda_1^{-1}$.  Using the
expression $\epsilon=\nu\langle(\nabla{\bbox v})^2\rangle$ for the
energy dissipation rate one can write the estimates
$\lambda_1\sim\sqrt{\epsilon/\nu}$ and $\Delta\sim\lambda_1$ for the
Lyapunov exponent and the dispersion.

\subsection{Single-Point Statistics}
\label{single}

In this subsection we examine the single-point statistics of the
conformation tensor ${\sl A}$ at $\lambda_1\tau<1$. As explained
above, most of the time ${\sl A}$ fluctuates near ${\sl A}_0$. We are
interested in large values ${\sl A}\gg {\sl A}_0$ because it is only
for large values of ${\sl A}$ that the polymers can possibly lead to
noticeable effects.  Large values are determined by the velocity
fluctuations such that the product $WW^{T}$ is anomalously large for a
long time. To find the tail of the PDF of ${\sl A}$ let us substitute
decomposition (\ref{product}) into Eq. (\ref{basic}). We obtain
\begin{eqnarray}&&
M^T{\sl A} M=\frac{2{\sl A}_0}{\tau}
\int_0^{\infty}dt\,\Lambda^2(t)
\exp\left[-\frac{2t}{\tau}\right]\,.
\nonumber\end{eqnarray}
where we used the $t$-independence of $M$ at large $t$. Under the condition
$e^{\rho_1}\gg e^{\rho_2}\gg e^{\rho_3}$ the tensor ${\sl A}$ is
uniaxial: 
\begin{eqnarray}&& 
{\sl A}_{ij}\approx{\sl T}{n}_i{n}_j\,.
\label{uniaxa}\end{eqnarray}
Here $\bbox{n}$ is a unit vector, $n_i=M_{i1}$, uniformly distributed
over a sphere, and ${\sl T}\equiv{\rm tr}\,{\sl A}$:
\begin{eqnarray}&&
{\sl T}\approx
\frac{2{\sl A}_0}{\tau}\int_0^{\infty}\!dt\, 
\exp\left[2\rho_1(t)-\frac{2t}{\tau}\right]\,.
\label{passiv7} \end{eqnarray}
Thus the matrix ${\sl A}$ is expressed via the scalar ${\sl T}$, which
is independent of $\rho_2$ and $\rho_3$. The statistics of ${\sl T}$
cannot be directly examined in terms of the single-time probability
distribution function (\ref{dis1}) because integral (\ref{passiv7})
involves different times. Nevertheless, it is possible to use PDF
(\ref{dis1}) to find the asymptotic behavior of the PDF of ${\sl T}$
at ${\sl T}\gg {\sl A}_0$. A rigorous derivation is presented in
Appendix \ref{integr} (cf. \cite{97CKV,98Cher}).
Below we use a simple semi-qualitative method.

For a typical fluctuation of the velocity, integral (\ref{passiv7})
converges at $t\sim\tau$, which gives ${\sl T}\sim {\sl A}_0$. To find
the probability distribution for large deviations of ${\sl T}$ one
should analyze rare events leading to a given value ${\sl T}\gg 
{\sl A}_0$, and find the event with the maximum probability. Let us
establish the structure of such fluctuations. It is obvious that
$\rho_1$ should initially grow faster than $t/\tau$ during some
interval of time. To ensure non-zero value of the probability of such
configuration, $\rho_1$ should then return to its average value,
$\lambda_1 t$. Since $\lambda_1<1/\tau$, the difference
$\rho_1(t)-t/\tau$ has a maximum at some time, $t=t_*$. At ${\sl T}\gg
{\sl A}_0$ the maximum is sharp and integral (\ref{passiv7}) is
determined by its vicinity. With the logarithmic accuracy
\begin{eqnarray}&&
\ln({\sl T}/{\sl A}_0)\approx 2\rho_1(t_*) -2 t_*/\tau \,.
\label{estim} \end{eqnarray}
The probability of the event is also determined by a vicinity of
$\rho_1(t_*)$ because it corresponds to the maximal deviation from the
average value of $\rho_1$. In accordance with Eq. (\ref{dis1}) it can
be estimated as
\begin{eqnarray}&&
\ln {\cal P}\approx -t_*
S_1\left(\frac{\rho_1(t_*)}{t_*}-\lambda_1\right) \,.
\nonumber \end{eqnarray}
Substituting here $\rho_1(t_*)$ expressed via ${\sl T}$ from Eq.
(\ref{estim}) and maximizing the result over $t_*$ we get the condition
\begin{eqnarray}&&
S_1\left(\beta+\frac{1}{\tau}-\lambda_1\right)-
\beta S_1'\left(\beta+\frac{1}{\tau}-\lambda_1\right)=0\,,
\label{passiv8}
\end{eqnarray}
where $\beta=(2t_*)^{-1}\ln({\sl T}/{\sl A}_0)$. Using the convexity
of $S_1$, one can show that Eq. (\ref{passiv8}) together with the
condition $\beta>0$ uniquely determines $\beta$. Then one finds
\begin{eqnarray}&&
\ln {\cal P}\approx-\frac{\alpha}{2}
\ln({\sl T}/{\sl A}_0) \,, \quad 
\alpha=S_1'\left(\beta+\frac{1}{\tau}-\lambda_1\right) \,.
\label{alpha} \end{eqnarray}
One can verify that the convexity of $S_1$ ensures the condition
$\alpha>0$ if $\lambda_1<1/\tau$. Expression (\ref{alpha}) determines
the probability density function of $\ln({\sl T}/{\sl A}_0)$. For the
PDF of $T$ we obtain
\begin{eqnarray}&&
{\cal P}(T) \sim \frac{{\sl A}_0^{\alpha/2}}{{\sl T}^{1+\alpha/2}}\,,
\label{powert} \end{eqnarray}
Since $A\propto R^2$, the power-law distribution of $A$ can be used to
obtain the power-law distribution of the molecular elongation, $R$
\cite{short}.

We see that the PDF is a power-law function with the exponent
$1+\alpha/2$ that can be expressed via the entropy function $S_1$. Since
the precise form of $S_1$ is generally unknown, it is impossible to find
the precise dependence of $\alpha$ on the parameters of the
flow. However, some general properties can still be inferred. As
$\lambda_1$ increases, i.e. when the Reynolds number increases,
$\alpha$ decreases and tends to zero when $\lambda_1\to1/\tau$. One
can easily establish the behavior of $\alpha$ in this region since
then the approximation (\ref{parab}) is correct. Substituting
Eq. (\ref{parab}) into Eqs.  (\ref{passiv8},\ref{alpha}) we obtain
\begin{eqnarray}&&
\alpha=\frac{2}{\Delta}
\left[\frac{1}{\tau}-\lambda_1\right]\,.
\nonumber \end{eqnarray} 
Note that the only characteristics of the flow entering this
expression are the average value of $\rho_1$ and its dispersion.

Power tail (\ref{powert}) means a slow decay of the probability
distribution of ${\sl T}={\rm tr}\,{\sl A}$, which results in infinite
values of its high moments. Namely, the moments $\langle{\sl T}^n
\rangle=\int d{\sl T}\,{\cal P}({\sl T}){\sl T}^n $ diverge if
$n>\alpha/2$. Moreover, the normalization integral $\int d{\sl
T}\,{\cal P}({\sl T})$ converges at large ${\sl T}$ only if
$\alpha>0$. Therefore an attempt to extend the passive consideration
to $\lambda_1>1/\tau$ leads to the divergence of the normalization
integral. It can be interpreted as the tendency of the polymer
molecules to be stretched, i.e. the coil-stretch transition, and the
breakdown of the passive approach. 

As we have seen, the $n$-th moment of the conformation tensor ${\sl
  A}$ formally diverges at $n\geq\alpha/2$. It signals the breakdown
of the passive approach, i.e. the main contribution to the diverging
moments comes from the configurations of the velocity such that the
feedback of the polymers on the flow cannot be disregarded. As
explained in the beginning of this section, the molecules can be
considered as passive at $\Pi\ll\nu\lambda_1$. An account of the back
reaction of the polymers on the flow leads to a much faster decrease
of the PDF of $\Pi$ at $\Pi\gtrsim\nu\lambda_1$, which ensures that
the moments have finite values (in the framework of the simple model
presented in Appendix \ref{model} one can find the precise form of the
PDF). Let us estimate the value of the diverging moments taking the
feedback into account. It is more convenient to discuss the moments of
$\Pi$, which are proportional to the moments of ${\sl A}$ (see Eq.
(\ref{stress})). If $\alpha<2$ then Eq. (\ref{powert}) modified by the
cutoff at $\Pi\sim \nu\lambda_1$ gives
\begin{eqnarray}&& 
\langle {\rm tr}\, \Pi \rangle \sim \Pi_0^{\alpha/2}
(\nu\lambda_1)^{1-\alpha/2} \,,
\label{avert} \end{eqnarray}
Note that $\langle{\rm tr}\,\Pi\rangle\gg\Pi_0$ because we assumed
that $\Pi_0\ll\nu\lambda_1$. The equations (\ref{dissip},\ref{avert})
show that the elastic contribution to the energy dissipation,
$\tau^{-1}\langle{\rm tr}\,\Pi\rangle$, can be estimated as $\tau^{-1}
\Pi_0^{\alpha/2} (\nu\lambda_1)^{1-\alpha/2}$.  It becomes comparable
to the viscous contribution, $\sim\nu\lambda_1^2$, just at the point
of the coil-stretch transition, where $\alpha=0$.

\subsection{Correlation Functions}
\label{corrf}

Here we investigate simultaneous many-point correlation functions of
${\sl A}$. Let us start with the two-point correlation function,
\begin{eqnarray}&&  
G_2({\bbox r})=\langle {\sl T}(t_0,\bbox{r}_1+\bbox{r})
{\sl T}(t_0,\bbox{r}_1)\rangle\,,
\label{pairdef} \end{eqnarray}
where ${\sl T}={\rm tr}\,{\sl A}$. The value of ${\sl A}$ at a given
point is determined by the Lagrangian trajectory arriving at this
point at $t=t_0$. Polymers separated by distances smaller than the
viscous scale, $\eta$, are stretched coherently, whereas at larger
separations the correlation is largely lost. Therefore $\eta$ is the
correlation length of ${\sl A}$. For distances $r\gtrsim \eta$ the
quantities ${\sl T}$ in (\ref{pairdef}) become weakly correlated and
the correlation function tends to the product of averages 
$\langle{\sl T}\rangle$. Non-trivial correlations occur at the
distances smaller than $\eta$. The correlation function is a
monotonically decreasing function of the distance $r$.

Note that $G_2$ cannot be calculated in the framework of the passive
approach if $\alpha<2$. Indeed, we can write $G_2\approx
\langle {\sl T}\rangle^2$ at $r\gg \eta$. Formal calculation of
$\langle{\sl T}\rangle$ using PDF (\ref{powert}) gives an infinite
result. The same is true for $G(0)=\langle {\sl T}^2\rangle$. Since
$G_2(r)$ is a decreasing function of $r$, it follows that $G_2(r)$ is
infinite in the framework of the passive approach. It means that the
main contribution to the correlation function comes from
configurations such that the back influence of the polymers on the
flow is not small. On the other hand, if $\alpha>4$ the calculation of
both $\langle{\sl T}\rangle^2$ and $\langle{\sl T}^2\rangle$ in the
passive framework gives a value of the order of ${\sl A}_0^2$. Hence
$G_2(r)\sim {\sl A}_0^2$ for all $r$, which means that the main
contribution is made by the configurations where the polymers are
close to the equilibrium state.

Let us consider the most interesting case $2<\alpha<4$. Then
$\langle{\sl T}\rangle\sim {\sl A}_0$ is finite (and small), whereas
$\langle{\sl T}^2\rangle$ is infinite if calculated using the
``passive'' PDF (\ref{powert}). It means that at small distances the
back reaction must be taken into account, whereas at larger distances
the passive approach works well. If we calculate the correlation
function in the framework of the passive approach, the result is valid
for distances larger than some characteristic scale of the back
reaction.

To calculate correlation function (\ref{pairdef}) one can
substitute expression (\ref{basic}) for ${\sl A}(\bbox r_1+\bbox r)$
and ${\sl A}(\bbox r_1)$ and then average over $\bbox r_1$ which can
be considered as averaging over space. The distance between Lagrangian
trajectories terminating at the points $\bbox r_1+\bbox r$ and $\bbox
r_1$ is an increasing function of $t$. It can be found from
Eq. (\ref{divl}):
\begin{eqnarray}&&
\delta\bbox x(t_0-t)=
W(t_0-t,t_0)\bbox r 
=W^{-1}(t_0,t_0-t)\bbox r \,.
\label{separ} \end{eqnarray}
Expression (\ref{separ}) is correct provided $|\delta\bbox
x|<\eta$. Under the same condition the matrices $W$ entering the
expressions (\ref{basic}) for ${\sl A}(\bbox r_1+\bbox r)$ and ${\sl
A}(\bbox r_1)$ are identical, as follows from Eq. (\ref{mmm1}).  Using
the decomposition (\ref{product}), rewritten as
$W^{-1}(t_0,t_0-t)=N^T\Lambda^{-1} M^T$, and the inequalities
$e^{\rho_1}\gg e^{\rho_2}\gg e^{\rho_3}$ we obtain $|\delta{\bbox
x}|\approx e^{-\rho_3} r$.

At $r\ll\eta$ we have $G_2\gg {\sl A}_0^2$, i.e. the main contribution
to $G_2$ is due to the rare events when the product $WW^T$ is
anomalously large during a long time. Then one can use the uniaxial
approximation (\ref{uniaxa}). The functions $\rho_1(t)$ in ${\sl
A}({\bbox r}_1+{\bbox r})$ and ${\sl A}({\bbox r}_1)$ are identical as
long as $|\delta{\bbox x}|<\eta$. When the separation $\delta{\bbox
x}$ becomes larger than $\eta$, the correlation between the Lagrangian
trajectories becomes weak. The contribution of this stage to $G_2$ is
given by the product of independent averages, $\langle{\sl T}\rangle^2
\sim {\sl A}_0^2$, and can be neglected. We conclude that the main
contribution to $G$ comes from times when $|\delta\bbox x|<\eta$.
Using the expression $|\delta\bbox x|\approx e^{-\rho_3}r$ we get
\begin{eqnarray}&& 
G\sim \left\langle\frac{{\sl A}_0^2}{\tau^2}
\left\{\int_0^sdt\,\exp
(2\rho_1-2t/\tau)\right\}^2\right\rangle \,,
\label{ggg} \end{eqnarray}
where $\rho_3(s)=\ln(r/\eta)$. Since both $\rho_1$ and $\rho_3$ enter
the integral, to evaluate $G_2$ one needs the joint PDF (\ref{pdf1}).

To ensure a large value of $G_2$, the function $\rho_1$ in
Eq. (\ref{ggg}) should first increase faster than $t/\tau$, and then
return to its average value $\lambda_1 t$. Thus $\rho_1-t/\tau$ should
have a maximum at $t_*<s$. A vicinity of $t=t_*$ makes the main
contribution to $G_2$. In the absence of the constraint $t_*<s$ the
value of $G_2$ grows exponentially as $t_*$ increases, which
corresponds to a formally infinite value of the second moment at
$\alpha<4$. Therefore the optimum is achieved at $t_*\approx s$. It
gives the estimate
\begin{eqnarray}&& 
G_2\sim {\sl A}_0^2\int d\rho_1\exp(4\rho_1-4s/\tau)
\nonumber\\&&\times
\exp\left\{-sS\left(\frac{\rho_1-\lambda_1s}{s},
\frac{\ln(r/\eta)-\lambda_3s}{s}\right)\right\} \,,
\label{ggg1} \end{eqnarray}
where $s$ is determined from the condition $\rho_3(s)=\ln(r/\eta)$.
The integral can be calculated in the saddle-point approximation with
the saddle-point $\rho_{1}^*\propto s$. Next, one should optimize over
$s$, which gives $s\propto\ln(r/\eta)$. The proportionality
coefficients depend on the form of $S$. The substitution of the
optimal values gives
\begin{eqnarray}&& 
G_2\sim {\sl A}_0^2 
\left({\eta}/{r}\right)^{\xi_2} \,.
\label{ggg2} \end{eqnarray}
The exponent $\xi_2$ in Eq. (\ref{ggg2}) can be found if the precise
form of the entropy function $S$ is known. We observe that $G_2\sim
{\sl A}_0^2$ if $r\sim \eta$. It is natural since $G_2$ at $r\sim
\eta$ can be estimated as $\langle {\sl T}\rangle^2$. We also see that
$G_2$ tends to infinity at $r\to0$. This corresponds to a formally
infinite value of the second moment.

All the conclusions concerning the pair correlation function of ${\sl
T}$ are valid for correlation functions of separate components of
${\sl A}$ too, which follows from the single-axis substitution
(\ref{uniaxa}). Indeed, ${\bbox n}$ is uniformly distributed over the
unit sphere, reduces correlation functions of ${\sl A}$ to correlation
functions of its trace ${\sl T}$.

Let us discuss the case $\alpha>4$. Then the main contribution to
$G_2(r)$ (\ref{pairdef}) at small $r$ is equal to the second moment
$\langle{\sl T}^2\rangle\sim {\sl A}_0^2$. One can examine the
$r$-dependent correction to the second moment $\langle{\sl
T}^2\rangle-G_2(r)$. It can be done as above. The correction behaves
as a positive power of $r$ at $r\ll\eta$.

The proposed scheme can be generalized to higher-order correlation
functions
\begin{eqnarray}&& 
G_n=\left\langle{\sl A}({\bbox r}_1)
\ldots{\sl A}({\bbox r}_n)\right\rangle
\label{ggnn}\end{eqnarray}
The behavior of $G_n$ is similar to that of $G_2$. If the moment
$\left\langle{\sl T}^n\right\rangle$ calculated with the PDF
(\ref{powert}) is infinite, the function $G_n$ is a scaling function
of the coordinates. The scaling exponent is negative, so the
correlation function formally diverges at small distances.  On the
other hand, if the moment $\langle{\sl T}^n\rangle$ is finite then the
difference $G_n-\langle{\sl A}^n\rangle$ scales with a positive
exponent and is thus a small correction to $\langle{\sl A}^n\rangle$.

Since the moments of $\Pi$ are finite, we can assert that
the growth of the correlation functions of $\Pi$ observed at fusing
points in Eq. (\ref{ggnn}) has to be saturated. For example, at
$2<\alpha<4$ the pair correlation function (\ref{pairdef}) saturates
at $G_2\sim{\sl A}_0^2\Pi_0^{-2}\langle\Pi^2\rangle \sim {\sl
A}_0^2\,\Pi_0^{\alpha/2-2}(\nu\lambda_1)^{2-\alpha/2}$. One can say
that the back reaction regularizes the correlation functions at small
scales.

\section{Strong Back Reaction}
\label{saturation}

Here we consider the dynamics of the polymer solutions above the
coil-stretch transition, when the Reynolds number exceeds a critical
value. Depending on the concentration of polymer molecules two
situations are possible. If the concentration is small, the elastic
stresses are small in comparison with the viscous stresses. Then the
polymers are stretched to their maximal elongation, $\Rm$, and the
properties of the fluid do not differ significantly from those of the
pure solvent. Below we consider the more interesting case where the
concentration of polymers is large enough, so that elastic stresses
can be larger than the viscous stresses. Then the feedback due to the
polymers substantially modifies the flow. The condition for the
existence of the back reaction regime is $\Pm\gg\nu/\tau$, where $\Pm$
is the maximal value of the elastic stress tensor. It can be expressed
in terms of microscopic parameters and the concentration of the
polymers as $\Pm=K_0n\varrho^{-1}\Rm^2$ (see subsect.\ref{conti}).
Using estimates for the microscopic parameters proposed in Ref.
\cite{Hinch77} one can rewrite the condition as $n\gg(R_0\Rm^2)^{-1}$.

Whereas in pure solvent typical gradients of the velocity grow
unlimited as the Reynolds number increases, in polymer solutions the
balance of inertial and elastic degrees of freedom fixes the
characteristic value of the gradient at $1/\tau$. Indeed, if the
instantaneous velocity gradient exceeds $1/\tau$, it extends the
polymers, so that the elastic stress grows and damps the gradient. On
the other hand, if the velocity gradient is much smaller than
$1/\tau$, the molecules contract and do not produce effect on the
flow. Then the velocity gradients tend to grow to the value
characteristic of the pure solvent, which is larger than $1/\tau$
above the transition. This explains the steady state realized above
the transition. We now establish some general properties of this
steady state.

Turbulence of Newtonian fluids can be characterized by two length
scales: the integral scale, $L$, and the dissipation scale, $\eta$.
Energy pumped at the integral scale cascades without dissipation from
larger to smaller eddies (coherent motions of the fluid) in the range
$\eta<r< L$, called the inertial interval. Velocity difference between
two points separated by the distance $r$ from the inertial range
diminishes slower than $r$, so that the characteristic value of the
velocity gradient at the scale $r$ grows downscales reaching a maximum
at $r\sim\eta$ \cite{Frisch}. If $V\tau/L\ll 1$, where $V$ is the
velocity at the integral scale, the gradient related to large eddies
is smaller than $\tau^{-1}$. Therefore, large eddies do not excite
polymers, which means that the elastic stress tensor is not correlated
at these scales. Since only coherent excitations of the elastic stress
tensor can influence the velocity, we conclude that the elasticity is
negligible for large eddies. The interaction of inertial and elastic
degrees of freedom becomes essential at the scale $r_*$, determined
from the condition $\nabla v\sim 1/\tau$. The fluctuations of $\Pi$
are correlated over the scale $r_*$. Because the value of the gradient
cannot exceed $1/\tau$, the velocity difference scales linearly with
$r$ at $r\lesssim r_*$, i.e. the flow is smooth. Near the coil-stretch
transition characteristic velocity gradient is determined by the
viscous scale and is of the order of $1/\tau$, hence $r_*\sim\eta$. As
the Reynolds number increases, velocity fluctuations increase, so that
the scale $r_*$ grows. Thus above the coil-stretch transition a new
scale, $r_*$, separating the inertial and viscoelastic intervals
arises. It is of the order of $\eta$ near the transition and grows as
the energy input increases.

Near the transition the viscous and elastic terms in
Eq. (\ref{intro1}) are of the same order, which gives
$\Pi\sim\nu/\tau$. For dilute solutions $\nu/\tau$ is much larger than
$\Pi_0$, therefore all the terms in the energy balance equation
(\ref{dissip}) are of the order of $\nu/\tau^2$ near the
transition. The energy pumping rate per unit mass, $\epsilon$, can be
estimated as $V^3/L$. Equating it to the dissipation rate, estimated
as $\nu/\tau^2$, one finds the estimate ${\rm
Re}_c=[L^2/(\nu\tau)]^{2/3}$ for the value of the Reynolds number at
the transition. As the energy input increases the energy dissipation
rate due to viscosity, $\nu(\nabla {\bbox v})^2$, remains of the order
of $\nu/\tau^2$. Therefore far above the transition the principal part
of the energy is dissipated by the polymer relaxation. Then the
viscous term in Eq. (\ref{dissip}) can be neglected and we obtain
\begin{eqnarray} && 
\langle{\rm tr}\,\Pi\rangle =\epsilon\tau \,.
\label{averpi} \end{eqnarray}
We conclude that the energy is dissipated mainly by the elastic
relaxation. Relation (\ref{averpi}) means that the typical value of
$\Pi$ grows as the energy input increases, which can be interpreted as
the increase in the effective (``elastic'') viscosity. It is defined
as the proportionality coefficient between the polymer stress tensor
$\Pi$ and the strain tensor $\nabla_iv_k+\nabla_kv_i$, which remains
of the order $1/\tau$. Using Eq. (\ref{averpi}) one can estimate the
ratio of the elastic term to the non-linear inertial term in Eq.
(\ref{intro1}) as $V\tau/L\ll 1$, which shows that the elasticity is
indeed negligible at large scales.

The strong interaction between the elastic and inertial degrees of
freedom imposes a restriction on the Lagrangian statistics of
velocity. To demonstrate it, observe that Eq. (\ref{intro2}) gets
simplified under the condition $\Pi\gg\Pi_0$ satisfied in the strong
back reaction regime. Neglecting the terms proportional to ${\sl A}_0$
and $\Pi_0$ in Eqs. (\ref{intro2},\ref{stress}) we obtain
\begin{eqnarray} &&
\partial_t {\Pi}_{ij}+(\bbox v\nabla){\Pi}_{ij}
={\Pi}_{kj}\nabla_k v_i+{\Pi}_{ik}\nabla_k v_j
-\frac{2}{\tau}{\Pi}_{ij} \,,
\label{intro5} \end{eqnarray}
Expressing the solution of Eq. (\ref{intro5}) in terms of the
Lagrangian quantities ${\bbox x}$ and $W$ introduced by Eqs. 
(\ref{lagrtr},\ref{mmm1}) we obtain
\begin{eqnarray}&&
{\Pi}(t,{\bbox r})=
W(t,0,{\bbox r}){\Pi}[0,\bbox x(0,{\bbox r})]
W^{T}(t,0,{\bbox r}) e^{-{2t}/{\tau}} \,.
\label{intro4} \end{eqnarray}
The Lagrangian correlation time at the scale $r_*$ is $\tau$. 
Therefore at $t\gg\tau$ the eigen-values of $W$ are strongly
separated so that $\Pi$ is uniaxial: 
\begin{equation}
\Pi_{ij}= n_i n_j\,{\rm tr}\,\Pi\,,
\label{uniax} \end{equation}
where $\bbox n$ is a unit vector. Then Eq. (\ref{intro4}) gives
\begin{eqnarray} &&
2\rho_1=\frac{2t}{\tau}
+\ln [{\rm tr}\, \Pi(t,{\bbox r})]
-\ln [{\rm tr}\, \Pi(0,{\bbox x}(0,{\bbox r}))] \,.
\label{stat} \end{eqnarray}

The stationarity of $\Pi$ implies that $\rho_1-t/\tau$ has a
stationary distribution. In particular, we conclude that the principal
Lyapunov exponent $\lambda_1$ of the flow is equal to $1/\tau$
exactly. The stationarity of $\rho_1-t/\tau$ is very different from
the situation for the Newtonian fluids, described by Eq. (\ref{dis1}).
The reason is the anticorrelations in the temporal dynamics of
$\sigma$ due to its interaction with $\Pi$ which were qualitatively
described in the beginning of the section. They lead to vanishing
dispersion $\Delta$ of $\rho_1-t/\tau$,
$\Delta=\int~dt\langle\langle\tilde\sigma_{11}(t)
\tilde\sigma_{11}(0)\rangle\rangle$, which is non-zero for Newtonian
turbulence.

Averaging Eq. (\ref{intro5}) one can obtain the exact relation
\begin{eqnarray} &&
\langle\Pi_{ik}({\bbox r})\nabla_kv_i({\bbox r})\rangle
=\frac{\langle {\rm tr}\Pi\rangle}{\tau}.
\label{sred} \end{eqnarray}
Consider now $\langle\Pi_{ik}({\bbox r})\nabla_kv_i({\bbox r'})
\rangle$ as a function of the separation $l=|{\bbox r}-{\bbox r'}|$.
Its value at $l=0$ is given by Eq. (\ref{sred}) and can be shown to be
much larger than the value at the pumping scale, $l\sim L$.  Indeed,
consider the correlation function averaged over a ball of size $L$
centered at ${\bbox r}$, i.e.  $\langle\Pi_{ik}({\bbox r})\int_V
d{\bbox r'} \nabla_kv_i({\bbox r'})\rangle/V$. The velocity gradient
averaged over the scale $L$ is determined by the external forces.
Using Eq. (\ref{averpi}) one can estimate the value of the averaged
correlation function as $\epsilon V\tau/L\ll\epsilon$. It follows
that $\langle\Pi_{ik} ({\bbox r})\nabla_kv_i({\bbox r'})\rangle$
decays at scales larger than $r_*$.  Below $r_*$ the fluctuations of
$\Pi$ and $\nabla v$ are strongly correlated. The decay of the
correlation function at $r_*<r<L$ can be used to derive the
Kolmogorov's four fifths law \cite{Frisch} at these scales. The latter
states that the third order longitudinal structure function is equal
to $-4\epsilon l/5$ in the inertial interval. All the above
conclusions are in agreement with the general picture presented in the
beginning of the section.

Expression (\ref{averpi}) gives the typical value of the stress
tensor. As we argued above, the fluctuations with $\Pi\gg\epsilon\tau$
relax rapidly due to the back reaction, which leads to a fast decrease
of the PDF of $\Pi$ at $\Pi\gg\epsilon\tau$. On the other hand, the
probability to have $\Pi\ll \epsilon\tau$ is also small. The rough
details of the behavior of the PDF can be understood on the basis of a
simple model presented in Appendix \ref{model}. The solution of the
model shows that the PDF of $\Pi$ has an exponential tail at large
values of $\Pi$ and power-law behavior at small values of $\Pi$. We
believe that a similar qualitative behavior is realized for the stress
described by Eqs. (\ref{intro1},\ref{intro5}). The model also
explicitly demonstrates the finite value of the Lagrangian correlation
time of $\Pi$ and $\nabla v$. This property holds despite a strong
modification of the Lagrangian dynamics due to the back reaction.

Note, that the concentration of the polymer molecules, $n$, does not
enter system of equations (\ref{intro1},\ref{intro5}). Therefore the
dynamics of polymer solutions with different values of $n$ will be
identical in the strong back reaction regime. Moreover, using the
equation $\partial_t n+({\bbox v}\nabla)n=0$ for the concentration, it
is possible to show that Eq. (\ref{intro5}) is also valid for
inhomogeneous in space $n$. Thus the hydrodynamic properties of
spatially inhomogeneous solutions do not differ from the homogeneous
ones. This assertion holds if local $n$ is large enough for $\Pm(n)$
to be larger than the local value of $\Pi$ prescribed by the dynamics.

The uniaxial form (\ref{uniax}) of the tensor $\Pi$ allows one to
rewrite Eqs. (\ref{intro1},\ref{intro5}) in the form similar to the
equations of the magnetic hydrodynamics. The field $\hat{\bbox
n}\,\sqrt{{\rm tr}\,\Pi}$ satisfies the induction equation with linear
damping. In addition one can show that the field
is solenoidal. This analogy helps understand the dynamics of
fluctuations at the scales $r\ll r_*$, which occur on the background
of the relatively slow stresses excited at $r\sim r_*$. These
small-scale fluctuations are elastic waves similar to the Alfven waves
propagating in the presence of a large-scale magnetic field in
magnetic hydrodynamics \cite{Kraichnan,MHD}. The dispersion relation
of the waves is $\omega=k\sqrt{{\rm tr}\,\Pi}$. Thus the velocity of
these waves is $\sqrt{{\rm tr}\,\Pi}$ which can be estimated as
$\sqrt{\epsilon\tau}$. There exist two mechanisms of the elastic waves
attenuation: polymer relaxation and viscous dissipation. The first
mechanism leads to the scale-independent attenuation $\tau^{-1}$,
which is smaller than the frequency, at $kr_*\gg1$. The second
mechanism produces the attenuation $\sim \nu k^2$ which is much
smaller than the frequency for $k\eta_*\ll 1$ where
$\eta_*=\nu(\epsilon\tau)^{-1/2}$. Thus the elastic waves are
well-defined in the interval $r_*^{-1}\lesssim k\lesssim\eta_*^{-1}$.

Our equations are valid as long as $\Pi\ll\Pm$. The relation
(\ref{averpi}) allows us to reformulate this condition as
$\epsilon\ll\Pm/\tau$. Another limitation of our scheme is related to
the inequality $R\ll r_*$, under which the flow is smooth at the scale
$R$. Using Eqs. (\ref{tlhs2}), (\ref{stress}), and (\ref{averpi}) one
can write the estimate $R^2\sim\varrho\epsilon\tau(K_0 n)^{-1}$ for
the typical size of a polymer molecule, $R$.

Let us estimate the parameters introduced above within the framework
of Kolmogorov's theory (K41) \cite{Kolmog}. Though the theory is,
rigorously, incorrect \cite{Frisch}, it is satisfactory for rough
estimates. The characteristic velocity difference, $\delta_r v$,
between two points separated by the distance $r$ from the inertial
interval is given in K41 by $(\epsilon r)^{1/3}$, where $\epsilon$ is
the energy input. Writing $|\nabla
v|\sim\delta_rv/r\sim\epsilon^{1/3}r^{-2/3}$, one finds
$r_*\sim\sqrt{\epsilon\tau^3}$. In the K41 theory the condition $R\ll
r_*$ can be rewritten as $\varrho(K_0 n\tau^2)^{-1}\ll 1$. Note that
in the framework of K41 theory the ratio $R/r_*$ is independent of the
Reynolds number.

Our analysis assumes that the characteristic size of the molecules,
$R$, is much smaller than their maximal size, $\Rm$. As ${\rm Re}$
increases, the typical elongations eventually become of the order of
$\Rm$, and further elongation becomes impossible. In this case the
molecules behave as hard rods, modifying the effective viscosity of
the fluid \cite{Batchelor}. Therefore at large enough ${\rm Re}$ we
return to the case of Newtonian fluid. However, this regime is
expected to be unstable because polymer molecules are intensively
destroyed by strong flows.

We have shown that in the steady state the velocity gradients in the
bulk do not exceed $\tau^{-1}$. Consider now the situation where the
boundary forces tend to produce gradients larger than $1/\tau$ at
$r\sim L$. Then the elastic reaction should lead to formation of a
boundary layer where the value of velocity gradient diminishes from
the value imposed by the forcing to the value $\tau^{-1}$ in the
bulk. Then $r_*\sim L$, i.e. the inertial range and energy cascade are
absent. This situation is similar to the elastic turbulence regime
\cite{00GS}.

Finally, let us consider the role of other modes of the polymer
molecules relaxation. They are characterized by the relaxation times
$\tau_i<\tau$. We have shown that the interaction of the fluid with
the principal relaxation mode fixes the value of the principal
Lyapunov exponent at $\lambda_1=\tau^{-1}$. The inequality
$\lambda_1\tau_i<1$ then implies that other modes are always only
weakly excited by the flow, so the interaction is fully determined by
the softest relaxation mode. We conclude that the equations
(\ref{intro1},\ref{intro5}) based on the single relaxation mode
approximation correctly describe the solution hydrodynamics above the
coil-stretch transition.

\section{Conclusion}
\label{conclusion}

We have examined properties of turbulence in dilute polymer solutions.
Our results support the theory of Lumley \cite{Lumley72} who argued
the existence of the coil-stretch transition in turbulent flows, which
occurs at a critical Reynolds number, ${\rm Re}_{\rm c}$. The polymer
molecules are typically weakly stretched, so that their elasticity
only weakly influences the flow in the regime realized below the
transition, at ${\rm Re}<{\rm Re}_{\rm c}$. At ${\rm Re}>{\rm Re}_{\rm
c}$ the polymer molecules are substantially stretched and strongly
modify the small-scale flow.

At ${\rm Re}<{\rm Re}_{\rm c}$ the polymer molecules are passively
advected and stretched by the flow. This regime occurs under the
condition $\lambda_1\tau<1$ where $\tau$ is the polymer relaxation
time and $\lambda_1$ is the principal Lyapunov exponent. The Lyapunov
exponent is defined as the logarithmic rate of the divergence of
nearby Lagrangian trajectories and can be estimated as the inverse
turnover time at the viscous scale of turbulence. The majority of the
molecules in this regime fluctuates near the equilibrium. There also
exists a small number of strongly elongated molecules which appear due
to rare large fluctuations in the rate of strain. Even though the
number of substantially elongated molecules is small, they may be
relevant in some situations due to the relatively slow power-law
decrease of the probability density function of elongations of
molecules (\ref{powert}).

In the second regime, at ${\rm Re}>{\rm Re}_{\rm c}$, most of the
molecules are substantially elongated. It leads to a strong
interaction between the elasticity and the flow, which modifies the
flow below the scale $r_*$. At $r\gtrsim r_*$ the properties of
turbulence are the same as in Newtonian fluids. The energy cascades
downscales from the pumping scale and dissipates due to polymer
relaxation at $r\sim r_*$. The scale can be considered as a new
dissipation scale. The flow is smooth at $r\lesssim r_*$ with the
Lyapunov exponent $\lambda_1$ fixed at the value $1/\tau$ by the
interaction.  

The smoothness of the flow at $r\lesssim r_*$ leads to the conclusion
that the velocity spectrum $E(k)$ decreases faster than $k^{-3}$ at 
$kr_*\gtrsim 1$. The precise form of $E(k)$ in this interval is
related to the elastic waves propagating at these scales. As both 
spectral transfer time and the decay time are scale-independent one 
can expect a power-law spectrum.

The properties of the polymer statistics near ${\rm Re}_{\rm c}$ were
examined numerically by Kronj\"ager and Eckhardt \cite{Eckhardt} in
the framework of Eqs. (\ref{intro2},\ref{intro1}). The results
indicate the power PDF tail for the polymer elongations at ${\rm
  Re}<{\rm Re}_{\rm c}$ and a substantial modification of the PDF at
${\rm Re}>{\rm Re}_{\rm c}$, in agreement with our results.

Let us discuss implications of our results for the drag reduction. A
description of the experimental situation can be found in the works
\cite{Virk,McComb,95GB}. It has been observed that the onset of the
drag reduction at increasing ${\rm Re}$ depends on the concentration
of the polymer molecules, whereas asymptotically the friction force
falls on a curve which is independent of the concentration. This curve
is usually referred to as MDR (maximum drag reduction) asymptote. A
discussion of the MDR can be found in the recent work \cite{Sreeni}. A
natural explanation of the $n$-independence of the MDR asymptote can
be formulated in the framework of Eqs. (\ref{intro1},\ref{intro5})
describing the strong back reaction regime. Indeed, the system
contains no $n$-dependent parameters. The $n$-dependence of the onset
can also be explained in our scheme. The drag force is formed in the
boundary layer which has a complicated structure \cite{Lumley69}.
Whereas gradients of the average velocity grow toward a wall, the
amplitude of the velocity fluctuations decreases. Therefore one can
expect that the polymer molecules are strongly extended in bulk and
weakly extended near the walls. Then the structure of the boundary
layer will be sensitive to the polymer concentration $n$. The
situation corresponds to the transient regime (which is sensitive to
the polymer concentration) from the Newtonian behavior to the MDR
asymptote.  The asymptote itself corresponds to the case when the
polymer molecules are strongly extended everywhere.

A striking property of polymer solutions is the so-called elastic
turbulence regime discovered by Groisman and Steinberg \cite{00GS}. It
is a chaotic state which is realized at small Reynolds numbers Re. Its
existence is made possible by the large value of Weissenberg number
${\rm Wi}=\tau V/L$ which implies a strong non-linearity of the
system. This state can also be investigated in the framework of our
scheme. The results will be published elsewhere.

Let us give numerical values of parameters appearing in our theory for
a typical experimental arrangement. For the number of monomers
$10^6-10^7$ one has $R_0\sim 10^{-5}$ $cm$, $\Rm\sim 10^{-2}$ $cm$,
and $\tau\sim 10^{-2}-10^{-1}$ $s$. Then using $n_{\rm c}\sim
(\Rm^2R_0)^{-1}$ one can obtain $0.1$ $ppm$ for the concentration,
$n_{\rm c}$, below which polymers have no effect on the flow. Let us
assume that the polymer concentration is $10$ $ppm$, the integral
length is $L\sim 10$ $cm$ and take the water viscosity, $\nu\sim
10^{-2}$ $cm^2/s$. Then the critical Reynolds number ${\rm Re}_{\rm
c}\sim (L^2/(\nu\tau))^{2/3}$ is of the order of $10^{4}$. Above the
coil-stretch transition the characteristic size of polymers is given
by $R\sim R_0 \sqrt{{\rm Re}^3\varrho\tau\nu^3/(k_{\rm B}TnL^4)}\sim
10^{-5}R_0{\rm Re}^{3/2}$. We obtain that $R\sim 10^2 R_0$ in the
vicinity of the transition, which is in agreement with the assumption
$R_0\ll R\ll \Rm$. Using the Kolmogorov's estimate $\eta\sim L{\rm
Re}^{-3/4}$ we find that at the transition $\eta\sim 10^{-2}$ $cm$,
which is of the order of $\Rm$. Thus the assumption $R\ll\eta$ is
satisfied. These estimates seem to fit the existing experimental data.

\acknowledgements

We thank M. Chertkov, B. Eckhardt, G. Falkovich, A. Groisman, I.
Kolokolov, V. Steinberg, M. Stepanov for valuable discussions. V. L.
acknowledges the grants of Israel Science Foundation and Minerva
Foundation. A.~ F. acknowledges the support of Arc-en-Ciel. E. B.
acknowledges the support by the National Science Foundation under
Grants No. 9971332, 9808595 and 0094569.

\appendix

\section{}
\label{integr}

Here we consider the statistical properties of the integral
\begin{eqnarray} &&
I=\int_0^{\infty} d t\,
\exp\left[\int_0^t d t'\,\xi(t')\right]\,, \quad
\label{intdef}\end{eqnarray}
where $\xi(t)$ is a random process with a finite correlation time
$\tau_{\xi}$ and a negative average $\xi_0<0$.  For the purpose it is
convenient to introduce the auxiliary object
\begin{eqnarray} &&
I(t)=\int_t^{\infty}dt'
\exp\left[\int_t^{t'} d t''\xi(t'')\right]. 
\nonumber \end{eqnarray}
Due to the stationarity of $\xi$ the statistics of $I(t)$, it is
independent of $t$. Separating the integration interval one finds the
relation
\begin{eqnarray}&&
I(t-\delta t)=I(t)\exp\left[\int_{t-\delta t}^t 
\xi(t')dt'\right]
\nonumber \\&&
+\int_{t-\delta t}^t dt'
\exp\left[\int_{t-\delta t}^{t'} \xi(t'')dt''\right].
\nonumber \end{eqnarray}
It follows that 
\begin{eqnarray} &&
\ln I(t-\delta t)=\ln I(t)+\int_{t-\delta t}^t \xi(t')dt'
\nonumber \\ &&
+\ln\left\{1+\frac{\int_{t-\delta t}^t dt'
\exp\left[\int_{t-\delta t}^{t'} \xi(t'')dt''\right]}{I(t)
\exp\left[\int_{t-\delta t}^t \xi(t')dt'\right]}\right\}\,.
\label{passive3} \end{eqnarray}
If $I(t)$ is sufficiently large, it is possible to neglect the last
term on the left-hand side. The exact condition is formulated below.
Observe that $\ln I(t)$ depends on the values of the noise at times
larger than $t$ so that the second term is independent of the first
provided $\delta t\gg\tau_{\xi}$. Therefore the probability
distribution function, $P(z)$, of $z(t-\delta t)\equiv\ln I(t-\delta
t)$ is given by the convolution of the distributions of $z(t)$ (which
is also equal to $P(z)$) and $\int_{t-\delta t}^t \xi(t')dt'$. The
latter has a probability function similar to the one (\ref{dis1}). We
thus obtain the integral equation
\begin{eqnarray} &&
P(z)=\int \frac{dz'}{\sqrt{2\pi t\Delta}} P(z') 
\exp\left[-\delta tS_{\xi}
\left(\frac{z-z'-\xi_0\delta t}
{\delta t}\right)\right],
\nonumber \end{eqnarray}
where $\xi_0=\langle\xi\rangle$ and $S_{\xi}$ is the entropy function 
characterizing $\xi(t)$. Since the kernel of the integral operator
depends on the difference $z-z'$ only, the solution of this equation
is $P(z)\propto \exp [-\alpha z]$. We obtain the following expression
for the tail of the PDF of $I\equiv e^z$ 
\begin{eqnarray} &&
P(I)\sim I^{-\alpha-1}.
\label{passiv4}\end{eqnarray}
Here $\alpha$ is determined from the condition
\begin{eqnarray} &&
\int\frac{dx}{\sqrt{2\pi t\Delta}} 
\exp\left[\alpha x-\delta t
S_\xi\left(\frac{x-\xi_0\delta t}
{\delta t}\right)\right]=1.
\label{passiv5} \end{eqnarray}
The solution $\alpha=0$ should be rejected. This integral can be
evaluated by the saddle-point method so that its value is determined
by the maximum of the exponent.  Taking its value at $x=\xi_0\delta t$
we conclude that in order to satisfy condition (\ref{passiv5}),
$\alpha$ and $\xi_0$ must have different signs. Thus if $\xi_0<0$,
then $\alpha>0$ and the normalization integral for PDF (\ref{passiv4})
converges at $I\to\infty$. On the other hand, if $\xi_0>0$, there
exists no well-defined distribution of $I$.

The two equations that implicitly define $\alpha$ are given by the
saddle-point condition
\begin{eqnarray} &&
\alpha=S'_\xi(\beta-\xi_0),
\nonumber \end{eqnarray}
where $\beta$ is the saddle-point value of $(z-z')/\delta t$, and the
condition 
\begin{eqnarray} &&
S_\xi(\beta-\xi_0)-\beta S'_\xi(\beta-\xi_0)=0.
\nonumber \end{eqnarray}
which follows from the condition that the integral (\ref{passiv5})
is equal to $1$. One should reject the formal solution $\beta=\xi_0$
of these equations corresponding to $\alpha=0$. It is easy to see that
$\beta$ is positive together with $\alpha$. Now we may formulate the
condition for the applicability of the power tail. It is valid
provided the third term in Eq. (\ref{passive3}) is indeed much
smaller than the second for those $I(t)$ that determine the PDF of
$I(t-\delta t)$. From $z-z'=\beta \delta t$ it follows that
$I(t)=I(t-\delta t)\exp[-\beta\delta t]$, so that we arrive at the
condition $I\gg \exp[\beta\delta t]/(\beta^2 \delta t)$ ($\xi$ is
estimated as $\beta$). The increment $\delta t$ constrained by the
condition $\delta t\gg\tau_{\xi}$. There are two cases to be
considered. If $\beta^{-1}\gg\tau_{\xi}$ one can use the choice
minimizing the above ratio $\delta t\sim\beta^{-1}$, so that Eq.
(\ref{passiv4}) is valid for $I\gg\beta^{-1}$. In the opposite case
$\beta^{-1}\leq\tau_{\xi}$ the power tail is valid for $I\gg
\exp[\beta\tau_{\xi}]/(\beta^2\tau_{\xi})$.  

At small $\xi_0$ one can use quadratic expansion for
$S_{\xi}(x)\approx x^2/(2\Delta)$ which gives
\begin{eqnarray} &&
\beta=-\xi_0,\ \ \ \alpha=-\frac{2}{\Delta}\xi_0.
\label{deltac} \end{eqnarray}
The entropy function becomes quadratic in the limit $\tau_{\xi}\to0$.
Thus the expression (\ref{deltac}) is valid for any $\xi_0$ in the
case of a short-correlated process $\xi$.

\section{Model of the back reaction}
\label{model}

Let us introduce a simple model that captures the most robust features
of the interaction between elastic and inertial degrees of freedom.
The model is formulated in terms of the system of equations for two
variables $\sigma$ and $x$. The equations are
\begin{eqnarray} &&
\frac{d x}{dt} = \sigma x+x_0 \,, 
\qquad \sigma= -x+\xi \,,
\label{mo2} \\ && 
\langle\xi\rangle=a \,, \qquad
\langle\!\langle\xi(t_1)\xi(t_2)\rangle\!\rangle
=2\delta(t_1-t_2) \,,
\label{mo3} \end{eqnarray}
where double brackets denote the irreducible part of the correlation
function.  The variable $\sigma$ models the rate-of-strain subtracted
by $1/\tau$ and $x$ models the elastic stress. The time derivative in
Eq.  (\ref{mo2}) represents the full derivative $\partial_t+{\bbox
v}\nabla$, i.e. we consider the Lagrangian dynamics. The product
$\sigma x$ represents the combination on the right-hand side of
Eq. (\ref{intro5}) and $x_0$ stands for ${\sl A}_0$ in Eq.
(\ref{intro2}). The second equation in system (\ref{mo2}) represents
the Navier-Stokes equation (\ref{intro1}). Since we consider dynamics
at the scale $r_*$, all the spatial derivatives can be estimated as
$1/r_*$. The term $-x$ describes the back reaction and $\xi$ models
the influence of larger scales, exciting the motion at $r\sim
r_*$. The average $a\equiv\langle\xi\rangle$ is negative below the
coil-stretch transition and positive above. If $a>0$ the term $x_0$ on
the right-hand side of the first equation in (\ref{mo2}) can be
disregarded.

Starting from the system of equations (\ref{mo2},\ref{mo3}) one can
derive the Fokker-Planck equation for the PDF of $x$:
\begin{eqnarray}&&     
\partial_t{\cal P}
=\partial_x\left[x\partial_x\left(x{\cal P}\right)\right]
-\partial_x\left[(x_0+(a-x)x){\cal P}\right]\,.
\label{mo5} \end{eqnarray} 
The normalized stationary solution of Eq. (\ref{mo5}) is
\begin{eqnarray} && 
{\cal P}_0(x)=\frac{1}{Z}x^{a-1}
\exp\left(-x-\frac{x_0}{x}\right)\,,
\label{statio} \end{eqnarray}
where $Z=2x_0^{a/2}K_a(2\sqrt{x_0})$ is the normalization factor. Here
$K_a$ is the MacDonald function. 

At $a<0$, which corresponds to system (\ref{intro2}-\ref{intro1})
below the transition, the properties of $\sigma$ are only
insignificantly modified by the interaction with the variable $x$.
For example, in the limit $x_0\ll 1$ one finds that
$\langle\sigma\rangle=a$, which is the same value as without the back
reaction. However, the back reaction is important for rare events when
$\xi$ is large. The interaction leads to the exponentially decaying
tail which makes all the moments of $x$ finite. This corresponds to
the picture presented in the main body of the text. Note that the
power tail is universal, i.e. force-independent, whereas the
exponential tail is an artifact of a zero correlation time of $\xi$
\cite{BF99}.

In the case $a>0$, i.e. above the transition, the limit $x_0\to 0$ is
regular. One obtains
\begin{eqnarray} &&
{\cal P}_0(x)=\frac{x^{a-1}\exp(-x)}{\Gamma(a)} \,,
\label{moo} \end{eqnarray}
where $\Gamma(x)$ is the Euler $\Gamma$-function. We observe that all
the positive moments of $x$ exist, because the back reaction stops the
growth of $x$. The average value of $x$ is given by $\langle x\rangle=
a$ so that $\langle\sigma\rangle=0$. These facts correspond to the
statements $\langle\Pi\rangle=\epsilon\tau$ and $\lambda_1=\tau^{-1}$
from Sec. \ref{saturation}.

Let us now investigate non-simultaneous correlation functions of $x$
above the transition, i.e. when $a>0$. Then we can assume $x_0=0$. We
need the Green function ${\cal G}(t,x,y)$ of Eq. (\ref{moo}), which
satisfies
\begin{eqnarray}&&
\partial_t{\cal G}\!-\!
\partial_x\left[x\partial_x
\left(x{\cal G}\right)\right]
+\partial_x\left[(a-x)x{\cal G}\right]
\!=\!\delta(t)\delta(x\!-\!y) \,,
\label{mo6} \end{eqnarray}   
with the condition ${\cal G}(t<0)=0$. The Green function should be
regular at $x=0$ and decrease faster than any power of $x$ at
$x\to\infty$. Using ${\cal G}$ one can find non-simultaneous
correlation functions of $x$ in the steady state:
\begin{eqnarray}&&
\left\langle f_1[x(t)] f_2[x(0)]\right\rangle
=\int d x\, d y\,{\cal P}_0(y)
{\cal G}(t,x,y) f_1(x) f_2(y)\,,
\nonumber \end{eqnarray}
where $f_1$ and $f_2$ are arbitrary functions and ${\cal P}_0(x)$ is
defined by Eq. (\ref{moo}). 

The Laplace transform of ${\cal G}(t,x,y)$ is
\begin{eqnarray} && 
G(\lambda,x,y)\equiv\int_0^\infty d t\,
\exp(-\lambda t){\cal G}(t,x,y)\,.
\label{laplace} \end{eqnarray}
It satisfies the equation
\begin{eqnarray}&&
\lambda G=\partial_x\left[x\partial_x
\left(x{G}\right)\right]
-\partial_x\left[(a-x)x{G}\right] 
+\delta(x-y) \,,
\label{mo} \end{eqnarray}
following from Eqs. (\ref{mo6},\ref{laplace}). The solution of
Eq. (\ref{mo}) can be expressed in terms of two independent solutions
of the homogeneous equation
\begin{eqnarray}&&
\lambda G_{1,2}=
\partial_x\left[x\partial_x
\left(x{G}_{1,2}\right)\right]
-\partial_x\left[(a-x)x{G}_{1,2}\right] \,.
\label{homo} \end{eqnarray}
Two independent solutions of Eq. (\ref{homo}) are
\begin{eqnarray}&& 
G_1=x^{k_1}e^{-x}F(-k_2-1, k_1-k_2+1,x)\,,
\label{generic1}\\&&
G_2=x^{k_2}e^{-x}\Psi(-k_1-1, k_2-k_1+1,x)\,,
\label{generic2} \end{eqnarray}
where
\begin{eqnarray} && 
\Psi(\alpha,\beta,x)=
\frac{\Gamma(1-\beta)}{\Gamma(\alpha-\beta+1)}
F(\alpha,\beta,x)\nonumber\\&&
+\frac{\Gamma(\beta-1)}{\Gamma(\alpha)}
x^{1-\beta}F(\alpha-\beta+1,2-\beta,x) \,.
\label{def} \end{eqnarray}
and $F(\alpha,\beta,x)$ is the confluent hypergeometric function
\cite{Brychkov}. The functions $G_1$ and $G_2$ satisfy the boundary
conditions at $x=0$ and $x\to\infty$ correspondingly. The parameters
$k_{1, 2}$ are
\begin{eqnarray}&&
k_{1, 2}=\frac{a-2\pm\sqrt{a^2+4\lambda}}{2} \,.
\label{mo9} \end{eqnarray} 
Matching functions (\ref{generic1},\ref{generic2}) at $x=y$ one
obtains the Green's function
\begin{eqnarray}&&
G(\lambda,x,y)=\frac{y^{1-a}e^{y}}{\sqrt{a^2+4\lambda}}
\frac{\Gamma(-1-k_2)}{\Gamma(k_1-k_2)}\bigl[
\theta(y-x)G_1(x)G_2(y)\nonumber\\&&
+\theta(x-y)G_1(y)G_2(x) \bigr]\,,
\label{more}\end{eqnarray}
where $\theta(x)$ is the Heaviside step function.

Expression (\ref{more}) allows one to establish analytical properties
of $G$ as a function of $\lambda$.  The function is analytic in the
half-plane ${\rm Re}\lambda>0$. There is a branch point at
$\lambda=-a^2/4$ with the cut going along the axis ${\rm Im}\lambda=0$
from the branch point to $-\infty$. In addition to the branch point
there also exist poles located at $\lambda=n(n-a)$ where $n$ is an
integer number (including zero), such that $n<a/2$. Thus the poles lie
between the origin and the branch point. One can easily find the pole
contribution to $G$ near $\lambda=0$ corresponding to $n=0$. Using
Eq. (\ref{more}) we obtain
\begin{eqnarray}&& 
G(\lambda,x,y)=
\frac{x^{a-1}\exp(-x)}{\Gamma(a)\lambda}+\dots \,,
\label{mo12} \end{eqnarray}
where dots mean terms regular in $\lambda$.     

The Green function ${\cal G}$ is expressed via its Laplace transform as
\begin{eqnarray} &&
{\cal G}(t,x,y)=\int_{A-i\infty}^{A+i\infty}
\frac{ d \lambda}{2\pi i}\exp(\lambda t) G(\lambda,x,y) \,,
\label{mo7} \end{eqnarray}
where the integration contour lies on the right of all singularities
of $G(\lambda)$. Shifting the integration contour in Eq. (\ref{mo7})
to the left we find that the first term in the right-hand side of
Eq. (\ref{irr}) is reproduced by the pole contribution (\ref{mo12})
and then
\begin{eqnarray}&&
{\cal G}_1(t,x,y)
=\int_{-\epsilon-i\infty}^{-\epsilon+i\infty}
\frac{ d \lambda}{2\pi i}\exp(\lambda t)G(\lambda,x,y) \,,
\label{mo13} \end{eqnarray}  
where $\epsilon>0$ is a small number and ${\cal G}_1$ is defined by
the relation
\begin{equation}
{\cal G}_1(t,x,y)={\cal G}_1(t,x,y)-{\cal P}_0(x)\,.
\label{irr} \end{equation}  
Shifting the integration contour in Eq. (\ref{mo13}) to the left we
encounter the branch point $\lambda=-a^2/4$ if $a<2$ or the pole
$\lambda=1-a$ if $a>2$. Therefore at large $t$ the $t$-dependence of
${\cal G}_1$ is $\exp(-a^2 t/4)$ if $a<2$ and $\exp[(1-a)t]$ if $a>2$.
Since
\begin{eqnarray} &&
\left\langle f_1[x(t)] f_2[x(0)]\right\rangle=
\langle f_1\rangle\, \langle f_2\rangle\nonumber\\&&
+\int\!\! d x\, d y\,{\cal P}_0(y) 
{\cal G}_1(t,x,y) f_1(x) f_2(y) \,.
\label{irre} \end{eqnarray}  
the correlations of $x$ decay exponentially in time.

One can write the asymptote of ${\cal G}_1(t,x,y)$ at large $t$ explicitly.
Let us first consider $a<2$. Then the integration contour in Eq.
(\ref{mo13}) can be deformed into a curve going around the cut
starting from $\lambda=-a^2/4$. Calculating the jump on the cut
we get the expression
\begin{eqnarray} &&
{\cal G}_1(t,x,y) \approx \frac{2\ln(y/x)c_1+c_1^2}
{16t^{3/2}\sqrt{\pi}} x^{a/2-1}y^{-a/2}e^{-a^2 t/4} \,.
\label{mo14} \end{eqnarray}
valid for small $x$, $y$. Here $c_1=2\psi(1)-\psi(-a/2)$ and $\psi(z)$
is the logarithmic derivative of the $\Gamma$-function. At $a>2$ with
the exponential accuracy the function ${\cal G}_1(t,x,y)$ is given by
the residue at $\lambda=1-a$:
\begin{eqnarray}&&
{\cal G}_1(t,x,y)\approx \frac{(a-1)F(-1, a-1, x)
F(-1, a-1, y)}{\Gamma(a-2)y}
\nonumber\\ &&
\times e^{(1-a)t}x^{a-2}e^{-x} \,.
\nonumber \end{eqnarray}
Finite correlation time of $\sigma$ follows from 
$\langle\sigma(t)\sigma(0)\rangle=-\partial_t^2\langle\ln
x(t)\ln x(0)\rangle$. The expectation value of
$x(t)=\exp[\int_0^t\sigma(t')dt']$ grows exponentially with time at
times much larger than the correlation time of $\sigma$ unless $\int
dt\langle\sigma(t)\sigma(0)\rangle=0$. Thus the back reaction stops
the growth of $x$ and gives rise to this peculiarity of the statistics
of $\sigma$.

\end{multicols}

\end{document}